\documentstyle[preprint,aps,prd,eqsecnum,tighten]{revtex}

\input boxedeps
\SetepsfEPSFSpecial
\HideDisplacementBoxes
\newcommand{\etal}{{\em et al.}}
\newcommand{\ie}{{\em i.e.}}

\newcommand{\eg}{{\em e.g.}}
\newcommand{\ibid}{\textit{ibid.}}
\newcommand{\tevcc}{\hbox{ TeV}\!/\!c^2}
\newcommand{\gevcc}{\hbox{ GeV}\!/\!c^2}

\newcommand{\gev}{\hbox{ GeV}}

\newcommand{\tev}{\hbox{ TeV}}

\newcommand{\fb}{\hbox{ fb}}

\newcommand{\lum}{\hbox{ cm}^{-2}\hbox{s}^{-1}}
\newcommand{\eqn}[1]{(\ref{#1})}
\def\slashiii#1{\setbox0=\hbox{$#1$}#1\hskip-\wd0\hbox to\wd0{\hss\sl/\/\hss}}

\newcommand{\rpv}{$\slashiii{R}$}
\newcommand{\rpar}{$R$-parity}
\newcommand{\mumu}{$\mu^{+}\mu^{-}$ collider}

\def\mand{\qquad {\rm and} \qquad}

\newcommand{\Qpl}[3]{{Phys. Lett.} {\bf #1,} #2 (19#3)}	       %
\newcommand{\Qprl}[3]{Phys. Rev. Lett. {\bf #1,} #2 (19#3)}    %
\newcommand{\prep}[3]{Phys. Rep. {\bf #1,} #2 (19#3)}		       %
\newcommand{\Qpr}[3]{{Phys. Rev. D}{\bf #1,} #2 (19#3)}		       %
\newcommand{\np}[3]{Nucl. Phys. {\bf #1,} #2 (19#3)}		       %
\newcommand{\npbps}[3]{Nucl. Phys. B (Proc. Suppl.) 
           {\bf #1,} #2 (19#3)}	                                       %
\newcommand{\ib}[3]{{\em ibid.\/} {\bf #1,} #2 (19#3)}		       %
\newcommand{\hepph}[1]{(electronic archive:	hep--ph/#1)}	       %

\begin{document}
\draft
\preprint{CERN-TH/2000/099  FERMILAB--PUB--00/114--T  MRI-PHY-P20000304   
          IITK-HEP-00-02 \phantom{}}

\title{Study of $R$-parity Violation at a $\mu p$ Collider}
\author{Marcela Carena,$^{1,\;2)}$\thanks{Email address: carena@fnal.gov}
	Debajyoti Choudhury,$^{3)}$\thanks{Email address: debchou@mri.ernet.in} 
	Chris Quigg,$^{1)}$\thanks{Email address: quigg@fnal.gov} and
	Sreerup Raychaudhuri$^{4)}$\thanks{Email address: sreerup@iitk.ac.in}
        }
\address{$^{1)}$ Theoretical Physics Department, 
		 Fermi National Accelerator Laboratory, 
		 P.O.\ Box 500, Batavia, Illinois 60510 USA \\
	 $^{2)}$ TH Division, CERN, CH-1211 Geneva 23, Switzerland \\
	 $^{3)}$ Mehta Research Institute, 
		 Chhatnag Road, Jhusi, Allahabad 211 019, India\\
	 $^{4)}$ Department of Physics, Indian Institute of Technology,
		 Kanpur 208 016, India
	}
\date{\today}
\maketitle
\begin{abstract}
   To explore the discovery possibilities of a high-energy muon-proton 
   collider, we examine signals that could arise from direct-channel 
   formation of supersymmetric-particle resonances through operators
   that do not respect $R$-parity.
\end{abstract}

% \narrowtext

\section{Introduction}\label{sec:intro}
Over the past several years, investigations into the feasibility of 
high-energy \mumu s have raised the tantalizing prospect that it may 
be possible to capture enormous numbers of muons ($\gtrsim 10^{20}$ 
per year) and store them in intense beams \cite{snowmass,status,rbp}. 
Although much development will be required,
muon colliders promise exquisite energy resolution in the 
few-hundred-GeV r\'{e}gime and may offer a path to the study of 
multi-TeV lepton-lepton interactions.  The physics opportunities have 
now been explored in some detail \cite{bbgh,CQmumu,SGSF}.
The technology that would make muon colliders a reality could also be 
applied to the production of intense neutrino sources from stored 
muon beams \cite{sgnu} that could be exploited for studies of deeply 
inelastic scattering or neutrino oscillations \cite{nubeams}.

If an energetic muon beam is stored in proximity to a high-energy 
proton beam, it is natural to consider the possibility of bringing 
them into collision.  A luminosity of $10^{32}\hbox{ -- }10^{33}\lum$ 
might be achieved in collisions of a stored $\mu^{\pm}$ beam with the 
1-TeV proton beam of the Fermilab Tevatron \cite{shiltsev}.  With a 
200-GeV muon beam, such a machine would have an impressive kinematic 
reach, with $\sqrt{s} \approx 0.9\tev$ and $Q^{2}_{\mathrm{max}} 
\approx 8 \times 10^{5}\gev^{2}$.  For comparison, DESY's $e^{\pm}p$ 
collider \textsc{hera} has operated recently with 27.5-GeV electrons 
on 920-GeV protons, for $\sqrt{s}\approx 0.32\tev$ and 
$Q^{2}_{\mathrm{max}}\approx 10^{5}\gev^{2}$.  The lifetime integrated 
luminosity of \textsc{hera} is projected as $1\fb^{-1}$.

Because of the high luminosity and the large kinematic reach, physics 
at high $Q^{2}$ is potentially very rich \cite{HSmumu,SRmumu}.  In one 
year of high-luminosity operation (\ie, at $10\fb^{-1}$), the $\mu p$ 
collider would yield about a million charged-current $\mu^{-}p \to 
\nu_{\mu}+\hbox{anything}$ events with $Q^{2}> 5000\gev^{2}$.  For 
comparison, the H1 detector at \textsc{hera} has until now recorded  
about 360 such $e^{\pm}p$ charged-current events, and can expect about 
6800 over \textsc{hera}'s lifetime \cite{cristid}.  The search 
for new phenomena, including leptoquarks and squarks produced in 
$R$-parity--violating interactions, would be greatly extended 
\cite{related}.

To quantify the discovery reach of a $\mu p$ collider, we explore 
supersymmetric processes mediated by \rpar-violating interactions 
in $\mu^{\pm}p$ collisions.  We find that direct-channel formation of 
squarks through \rpv\ couplings with ordinary particles can produce 
sharp peaks in the invariant-mass distribution and dramatic 
enhancements in the $Q^{2}$ distribution.  The search for these 
effects has the potential to significantly increase experimental 
sensitivity to \rpv\ couplings.  This work complements an earlier 
study of the manifestations of \rpar\ violation in ultrahigh-energy 
neutrino interactions \cite{Carena:1998gd}.

\section{$R$-parity and supersymmetry}
Electroweak gauge invariance forbids terms in the standard-model 
Lagrangian that change either baryon number or lepton number.  Such 
terms are allowed in the most general supersymmetric (SUSY) extension 
of the standard model \cite{why}, but they may lead to an 
unacceptably short proton lifetime.  One way to evade the proton-decay 
problem is to impose a discrete symmetry called 
\rpar, which implies a conserved multiplicative quantum number, 
$R\equiv (-1)^{3B+L+2S}$, where $B$ is baryon number, $L$ is lepton 
number, and $S$ is spin \cite{ff}.  All ordinary particles are 
\rpar\ even, while all superpartners are \rpar\ odd.  If \rpar\ is 
conserved, superpartners must be produced in pairs and the lightest 
superpartner, or LSP, is absolutely stable.

Imposing \rpar\ invariance on the SUSY Lagrangian is an \textit{ad 
hoc} remedy not derived from any known fundamental principle.  For 
this reason alone, it is of interest to consider an \rpar--violating 
extension of the minimal supersymmetric standard model.  
%%%%%%%%%%%%%%%%%%%%%%%%%%%%%%%%%%%%%%%%%%%%%%%%%%%%%%%%%%%%%%%%%%%%%%%%%%%%%%%
%                                                                             %
%   What is more, \rpv\ interactions can improve the agreement between        %
%   theory and precision electroweak measurements (\eg, $Z^{0} \to            %
%   b\bar{b}, c\bar{c}$), and also offer ready explanations \cite{flocons}    %
%   for experimental anomalies such as the high-$Q^{2}$ excess reported at    %
%   HERA \cite{H1,ZEUS}.                                                      %
%                                                                             %
%%%%%%%%%%%%%%%%%%%%%%%%%%%%%%%%%%%%%%%%%%%%%%%%%%%%%%%%%%%%%%%%%%%%%%%%%%%%%%%

The most general \rpv\ terms in the superpotential consistent with 
Lorentz invariance, gauge symmetry, and supersymmetry 
are\footnote{We suppress here the SU(2)$_{\mathrm{L}}$ and SU(3)$_c$ indices.  
      Symmetry under SU(2)$_{\mathrm{L}}$ implies that the first term is 
      antisymmetric under $i\leftrightarrow j$, while SU(3)$_c$ symmetry 
      dictates that the third term is antisymmetric under 
      $j\leftrightarrow k$.  We neglect bilinear terms that mix lepton 
      and Higgs superfields \protect\cite{hallsuz}.  
      Discussions of the phenomenological implications of such terms can 
      be found in the literature~\protect\cite{BILIN}.  }
\begin{equation}
    W_{\not{R}}  =  \lambda_{ijk} L^i L^j \bar{E}^k 
+\lambda^{\prime}_{ijk} L^i Q^j \bar{D}^k 
   + \lambda^{''}_{ijk} \bar{U}^i \bar{D}^j \bar{D}^k 
\label{eq:superpot}
\end{equation}
where $i,j,k$  are generation indices,
$L^i \ni (\nu^{i},e^{\,i})_{\mathrm{L}}$ and  
$Q^i \ni (u^{i},d^{\,i})_{\mathrm{L}}$ are the left-chiral
superfields, and 
$E^i \ni e^{\,i}_{\mathrm{R}}$, $D^i \ni d^{\,i}_{\mathrm{R}}$, and 
$U^i \ni u^{i}_{\mathrm{R}}$ 
are the right-chiral superfields, respectively.
 The Yukawa couplings 
$\lambda_{ijk}$, $\lambda^{\prime}_{ijk}$, and $\lambda^{''}_{ijk}$ 
are \textit{a priori} arbitrary, so the \rpv\ 
superpotential \eqn{eq:superpot} introduces 45 free 
parameters.

The $LLE$ and $LQD$ terms change lepton number, whereas the $UDD$ term 
changes baryon number.  Since we wish to explore \rpv\ effects in $\mu 
p$ collisions, we shall explicitly forbid the $UDD$ interactions 
\cite{leptopar} as the most economical way to avoid unacceptably rapid 
proton decay.  When we expand the superfield components in 
\eqn{eq:superpot}, we obtain the interaction Lagrangian that 
contributes to $\mu^{\pm}p$ interactions,
\begin{equation}
    {\mathcal{L}}_{LQD} = \lambda^{\prime}_{ijk} \left\{
 \tilde{\nu}_{\mathrm{L}}^i d_{\mathrm{L}}^j \bar{d}^k_{\mathrm{R}} -
  \tilde{e}_{\mathrm{L}}^i u_{\mathrm{L}}^j \bar{d}^k_{\mathrm{R}}  
  + \tilde{d}_{\mathrm{L}}^j \nu_{\mathrm{L}}^i 
  \bar{d}^k_{\mathrm{R}} 
  -   \tilde{u}_{\mathrm{L}}^j e_{\mathrm{L}}^i \bar{d}^k_{\mathrm{R}}  
  + \tilde{d}_{\mathrm{R}}^{kc} \nu_{\mathrm{L}}^i d^j_{\mathrm{L}} -  
  \tilde{d}_{\mathrm{R}}^{kc} e_{\mathrm{L}}^i u^j_{\mathrm{L}}  \right\} + 
  \mathrm{H.c.}
\label{eq:LQD}
\end{equation}
The \rpv\ couplings in \eqn{eq:LQD} modify 
supersymmetric 
phenomenology in several important ways: processes that change lepton 
number are allowed, superpartners can be produced singly, and the 
LSP---now unstable against decay into ordinary particles---is no 
longer constrained by potential cosmological embarrassments
to be a neutral color singlet.

The remarkable agreement between present data and standard-model (SM)
expectations implies very restrictive bounds on the strength of many \rpv\ 
operators \cite{BGH,gb_dc,goity,gautam,herbi}.  Experimental limits on $LQD$
couplings of muons with the first-generation quarks 
found in nucleon targets are not terribly restrictive.  Muon couplings to 
second- and third-generation quarks are still less constrained.  
In Table \ref{tab:lambdas}, we summarize the constraints on the \rpv\ 
Yukawa couplings that are relevant for $\mu^{\pm}p$ collisions, for 
the case of a 200-GeV$\!/\!c^{2}$ sfermion.  In each example we 
consider, we shall assume that only one \rpv\ coupling can be sizeable 
at a time \cite{products}.

\rpar--violating interactions have been looked for in many 
experiments.  At the Tevatron Collider, for example, squark pair 
production and subsequent decays through an \rpv\ coupling could lead 
to an excess of events with a dilepton pair along with 
jets~\cite{Tev_dilep}.  More interestingly, (Majorana) gluinos can 
decay through both squarks and antisquarks to produce like-sign 
dileptons \cite{Tev_likesign}.  Until superpartners are discovered, 
all such analyses, of necessity, rest on \textit{ad hoc} assumptions 
about the spectrum.  Moreover, they cannot determine the {\em 
strength} of the \rpv\ coupling.  The last criticism does not apply to 
the HERA experiments \cite{H1_LQ} or to Drell-Yan processes at the 
Tevatron Collider \cite{Tev_DY}.

\section{$R$-parity--violating signals at a muon-proton collider}\
\label{sec:signals}
\subsection{General observations}\label{subsec:gen}
The best signature for \rpv\ $LQD$ couplings in $\mu p$ collisions is 
the resonant production of a squark.  If the squark decays through the 
same \rpar--violating coupling that produced it, this process can 
modify the cross section for deeply inelastic scattering.  If the 
squark also decays with significant probability through 
\rpar--conserving interactions, distinctive signals may arise from the 
cascade decays of the squark through gaugino channels.  In this 
article we analyze in detail only the first alternative, for which the 
signals consist of a single hard jet recoiling against a hard, 
isolated muon or neutrino.\footnote{A further possibility for 
charged-current events is the decay of a squark into a quark, which 
materializes as a jet, and the lightest neutralino, which subsequently 
decays outside the detector.  This occurs naturally for a 
Higgsino-dominated LSP.} The analysis is parallel to the case of 
leptoquark production and decay in $\mu p$ colliders.

In $\mu^{+}p$ scattering, the most important elementary process is
\begin{equation}
	\mu^{+}d \to \tilde{u}_{L}^{j} \to \mu^{+}d ,
	\label{eq:mupnc}
\end{equation}
the interaction of a $\mu^{+}$ with a valence down quark through the 
$\lambda^{\prime}_{2j1}$ coupling, leading to a $\mu^{+}+\hbox{jet}$ 
final state.  Suppressed 
processes\footnote{In this section, we remark only on the resonant processes.
	However, all our quantitative studies include the full set of Feynman
	diagrams.}
involving sea quarks are
\begin{equation}
	\mu^{+}s \to \tilde{u}_{L}^{j} \to \mu^{+}s ,
	\label{eq:mupsnc}
\end{equation}
through a $\lambda^{\prime}_{2j2}$ coupling, which leads to a 
$\mu^{+}+\hbox{jet}$ final state, and 
\begin{equation}
	\mu^{+}\bar{u} \to \tilde{d}_{R}^{kc} \to 
	\left\{
		\begin{array}{c}
		\mu^{+}\bar{u} \\
		\bar{\nu}_{\mu}\bar{d}	
		\end{array} 
	\right. ,
		\label{eq:mupubar}
\end{equation}
through a $\lambda^{\prime}_{21k}$ coupling, which leads with equal 
probability to a ($\mu^{+}+\hbox{jet}$) or (jet + missing energy) signature.

In $\mu^{-}p$ scattering, the most important elementary process is
\begin{equation}
	\mu^{-}u \to \tilde{d}_{R}^{k} \to 
	\left\{
		\begin{array}{c}
		\mu^{-}u \\
		\nu_{\mu}d	
		\end{array} 
	\right. ,
		\label{eq:mumu}
\end{equation}
the interaction of a $\mu^{-}$ with a valence up quark through the 
$\lambda^{\prime}_{21k}$ coupling, which leads with equal 
probability to $\mu^{-}+\hbox{jet}$ or jet + missing energy signatures.
The interactions with light sea quarks are
\begin{equation}
	\mu^{-}\bar{d} \to \tilde{u}_{L}^{jc} \to 
	\mu^{-}\bar{d}  
	\label{eq:mumdbar}
\end{equation}
through the $\lambda^{\prime}_{2j1}$ coupling, and 
\begin{equation}
	\mu^{-}\bar{s} \to \tilde{u}_{L}^{jc} \to 
	\mu^{-}\bar{s}  
	\label{eq:mumsbar}
\end{equation}
through the $\lambda^{\prime}_{2j2}$ coupling.  Both of these suppressed 
reactions lead to $\mu^{-}+\hbox{jet}$ signatures.

The largest cross sections---and the most promising 
signals---should arise from interactions with valence quarks, the 
reactions \eqn{eq:mupnc} and \eqn{eq:mumu}.  We will analyze in detail 
the neutral-current reaction\footnote{The neutral-current reaction 
$\mu^{-}u \to \tilde{d}_{R}^{k} \to \mu^{-}u$ occurs with similar 
cross section; the valence up-quark density is roughly twice the 
valence down-quark density, but the $\tilde{d}_{R}^{k}$ branching 
fractions into $\mu^{-}u$ or $\nu_{\mu}d$ are one-half, whereas the 
$\tilde{u}_{L}^{k}$ branching fraction into $\mu^{+}d$ is unity.} 
$\mu^{+}d \to \tilde{u}_{L}^{j} \to \mu^{+}d$, and the charged-current 
reaction $\mu^{-}u \to \tilde{d}_{R}^{k} \to \nu_{\mu}d$.  We consider 
muon beams of 50 and $200\gev$ colliding with a 1-TeV proton beam.

A 1-TeV proton beam can be regarded as a broad-band, unseparated beam 
of quarks, antiquarks, and gluons, with energies typically in the 
range $(0,350)\gev$.  Accordingly, the c.m.\ frame for collisions of 
valence quarks with 200-GeV muons approximately coincides with the 
laboratory (collider) frame.  The appropriate detector is therefore symmetric, 
similar in concept to the current generation of general-purpose 
detectors at the Tevatron proton-antiproton collider, but emphasizing 
the detection and rejection of muons.  Background from the muon halo 
around the muon beamline probably requires cutting out a cone of 
about $\pm 10^{\circ}$ around the beamline.  The study of low-$x$ collisions 
appears very 
difficult because of the asymmetric kinematics and the angular cutoff.

\subsection{$\mu \lowercase{p}$ collider kinematics}\label{subsec:cine}
The natural kinematic observables for the inclusive neutral-current 
reaction $\mu p \to \mu+\hbox{anything}$ are the energy and 
angle of the outgoing muon in the collider frame, $E_{\mu}$ and 
$\theta_{\mu}$.  These quantities are not affected by hadronic 
fragmentation, and so distributions can be calculated reliably using 
parton-level Monte Carlo simulations.
At high energies, where we may 
safely neglect the muon and proton masses, 
we denote the incoming proton 
momentum in the collider frame by
\begin{equation}
	P = (E_{p};0,0,E_{p}) ,
	\label{eq:proton}
\end{equation}
so that it defines the positive $z$-axis.  The incoming muon momentum 
is
\begin{equation}
	p = (E_{\mu}^{0};0,0,-E_{\mu}^{0}),
	\label{eq:muin}
\end{equation}
and the outgoing muon momentum is
\begin{equation}
p^{\prime} = (E_{\mu}; E_{\mu}\sin\theta_{\mu},0,E_{\mu}\cos\theta_{\mu}).
	\label{eq:muout}
\end{equation}
We follow the conventions evolved for the analysis of $ep$ collisions 
at \textsc{hera}, in which the angle of the outgoing charged lepton is 
measured with respect to the \textit{proton} direction.  The momentum 
transfer is defined as $q \equiv p - p^{\prime}$, the square of the 
c.m.\ energy is
\begin{equation}
	s = 2 p \cdot P = 4E_{\mu}^{0}E_{p}.
	\label{eq:sdef}
\end{equation}

The invariant momentum transfer variable $Q^{2}\equiv -q^{2}$ and the 
Bjorken scaling variables $x$ and $y$ can be expressed in terms of 
the muon energy and angle in the collider frame as
\begin{eqnarray}
	Q_{\mu}^{2} & = & \displaystyle 
		 4 E_{\mu}^{0} E_{\mu} \cos^{2}\frac{\theta_{\mu}}{2} \ , 
	\label{eq:Q2_y_x_def}\\[1.5ex]
	y_{\mu} & \equiv & \displaystyle
		\frac{q \cdot P}{p \cdot P} 
	         =  \displaystyle 1 - 
	\frac{E_{\mu}}{E_{\mu}^{0}}\sin^{2}\frac{\theta_{\mu}}{2} ,
	\label{eq:ydef}
		\\[1.5ex]
	x_{\mu} & \equiv & 
	\displaystyle \frac{Q_{\mu}^{2}}{2q \cdot P}  =  \displaystyle
	\frac{Q_{\mu}^{2}}{2y_{\mu}p \cdot P} = 
	\frac{Q_{\mu}^{2}}{4y_{\mu}E_{\mu}^{0}E_{p}}  .\label{eq:xdef}
\end{eqnarray}
Combining eqns. \eqn{eq:Q2_y_x_def} and \eqn{eq:ydef}, we obtain the useful 
expression
\begin{equation}
	Q_{\mu}^{2}= \frac{(E_{\mu}\sin\theta_{\mu})^{2}}{1-y_{\mu}} = 
	\frac{p_{\perp\mu}^{2}}{1-y_{\mu}}  .
	\label{eq:Q2def2}
\end{equation}
The invariant-mass-squared of the outgoing hadronic system is
\begin{equation}
	M_{\mu}^{2} = x_{\mu}s = \frac{Q_{\mu}^{2}}{y_{\mu}}.
	\label{eq:Mdef}
\end{equation}
Following the example of the \textsc{hera} terminology, we call this technique 
for determining the kinematic invariants the \textit{muon method.}

For the inclusive charged-current reaction $\mu p \to 
\nu_{\mu}+\hbox{anything}$, we reconstruct the kinematic invariants 
using a parton-level modification of the Blondel-Jacquet technique 
employed in $ep$ experiments at \textsc{hera} \cite{bljq}.  Let
\begin{equation}
	p_{H} \equiv \sum_{h} (E_{h},p_{xh},p_{yh},p_{zh})
	\label{eq:pHdef}
\end{equation}
be the four-momentum of the outgoing hadronic system, summed over all 
hadronic clusters $h$.  Then we may write the momentum transfer as $q = 
p_{H}-P$, whereupon
\begin{equation}
	y_{H} = \frac{(p_{H}-P)\cdot P}{p \cdot P} = 
	\frac{\sum_{h}(E_{h}-p_{zh})}{2E_{\mu}^{0}}  .
	\label{eq:yHdef}
\end{equation}
Expressing \eqn{eq:Q2def2} in terms of hadronic variables, we have
\begin{equation}
	Q_{H}^{2} = \frac{\vec{p}_{\perp H}^{\;2}}{1-y_{H}} = 
	\frac{(\sum_{h}\vec{p}_{\perp h})^{2}}{1-y_{H}}  ,
	\label{eq:QHdef}
\end{equation}
where $\vec{p}_{\perp H}$ is the total transverse momentum of the 
hadronic flow and $\vec{p}_{\perp h}$ is the transverse momentum of 
hadron $h$.  We determine $x_{H}$ from the condition \eqn{eq:xdef} and 
express the invariant-mass-squared of the outgoing hadronic system as
\begin{equation}
	M_{H}^{2} = \frac{Q_{H}^{2}}{y_{H}}  .
	\label{eq:MHdef}
\end{equation}

For the parton-level simulation we present in Sec.\ 
\ref{subsec:CCsigs}, it is appropriate to use the kinematic variables 
of the struck parton that gives rise to the hard jet, instead of 
summing over hadron energies.  The kinematic invariants determined by 
this \textit{parton method} are then
\begin{equation} 
   y_{J}     =  \displaystyle \frac{E_{J}-p_{zJ}}{2E_{\mu}^{0}} \ , \quad
   Q_{J}^{2} = \displaystyle \frac{\vec{p}_{\perp J}^{\;2}}{1-y_{J}}  \ ,  
   \mand
   M_{J}^{2} = \displaystyle \frac{Q_{J}^{2}}{y_{J}} \ .
	 \label{eq:Jdef}
\end{equation}
% \begin{equation}
% 	y_{J} = \frac{E_{J}-p_{zJ}}{2E_{\mu}^{0}}  ,
% 	\label{eq:yJdef}
% \end{equation}
% \begin{equation}
%	Q_{J}^{2} = \frac{\vec{p}_{\perp J}^{\;2}}{1-y_{J}}  ,
%	\label{eq:Q2Jdef}
% \end{equation}
% and 
% \begin{equation}
%	M_{J}^{2} = \frac{Q_{J}^{2}}{y_{J}}  .
%	 \label{eq:MJdef}
%\end{equation}

In the analysis that follows, we use a parton-level Monte Carlo event 
generator \cite{MC} to compute tree-level 
cross sections.
%%%%%%%%%%%%%%%%%%%%%%%%%%%%%%%%%%%%%%%%%%%%%%%%%%%%%%%%%%%%%%%%%%%%%%%%%%
%                                                                        %
%   \footnote{It has been shown \textbf{[REFERENCE NEEDED]}              %
%   that next-to-leading-order calculations tend to enhance the cross    %
%   sections for both signal and background by a factor of 1.3 -- 1.5    %
%   \dots  THIS SEEMS WOOLY TO ME.  DO WE REALLY NEED THIS REMARK?}      %
%                                                                        %
%%%%%%%%%%%%%%%%%%%%%%%%%%%%%%%%%%%%%%%%%%%%%%%%%%%%%%%%%%%%%%%%%%%%%%%%%%
We treat the partons as observables: the struck quark is identified 
with a jet of the same four-momentum.  The numerical results we 
present are based on the CTEQ4-M parton distribution functions 
\cite{CTEQ4}. However, we have verified that our results are not 
very sensitive to the exact choice of the parton distributions 
and hence stable under changes to other standard
parametrizations.

\subsection{Neutral-current interactions}\label{subsec:NCsigs}
The neutral-current reaction $\mu^{+} p \to 
\mu^{+}+\hbox{anything}$ is mediated by $t$-channel $\gamma$ and 
$Z^{0}$ exchange in the standard electroweak theory.
The inclusive cross section is given by
\begin{equation}
  \sigma(\mu^+p \to \mu^+ X) 
     = \sum_{q} \int_0^1 {\rm d} x \: f_{q}(x, Q^2) \: 
                  \hat{\sigma}(\mu^+q \to \mu^+q) \ ,
    \label{eq:siginc}
\end{equation}
where $\hat{\sigma}(\mu^+q \to \mu^+q)$ is the elementary 
cross section for the $\mu^+$ to scatter off quark $q$ with momentum
fraction $x$. The flux of quarks $q$ in the proton is denoted by the 
parton distribution function $f_{q}(x, Q^2)$.
Identifying the subprocess Mandelstam variables 
with those used so far ({\em viz.} $\hat{s} = 4xE_{\mu}^{0} E_p$ and 
$\hat t \equiv - Q^2$), the parton-level cross sections are given by 
\begin{equation} \displaystyle
\frac{{\rm d} \hat{\sigma}}{{\rm d} \hat t} (\mu^- + q \to \mu^- + q) = 
	    \frac{1}{16 \pi \hat s^2}
             \Bigg\{
		\hat s^2 \Big[ \left| A_{LL} \right|^2 + \left| A_{RR} \right|^2
                 \Big]
              + \hat u^2 \Big[ \left| A_{LR} \right|^2 + \left| A_{RL} \right|^2
                 \Big]
	     \Bigg\} \ .
	\label{nc:cross-sec}
\end{equation}
The helicity amplitudes $A_{ab}$  assume very simple forms in the case of 
massless fermions. For example, the standard-model amplitudes can be 
expressed in the compact form
\begin{equation}
\begin{array}{rclcrcl}
A^{\mathrm{SM}}_{LL}(q) & = & \displaystyle
             e^2 \sum_{i = \gamma,Z} \frac{ L_i(\mu) L_i(q)}{ \hat t - m^2_i}
      & \qquad &
A^{\mathrm{SM}}_{LR}(q) & = & \displaystyle
             e^2 \sum_{i = \gamma,Z} \frac{ L_i(\mu) R_i(q)}{ \hat t - m^2_i}
        \\[3ex]                           
A^{\mathrm{SM}}_{RL}(q) & = & \displaystyle
             e^2 \sum_{i = \gamma,Z} \frac{ R_i(\mu) L_i(q)}{ \hat t - m^2_i}
        & & 
A^{\mathrm{SM}}_{RR}(q) & = & \displaystyle
             e^2 \sum_{i = \gamma,Z} \frac{ R_i(\mu) R_i(q)}{ \hat t - m^2_i}
        \\[5ex]                           
L_\gamma(f) & = & \displaystyle
             e_f 
         & & 
R_\gamma(f) & = & \displaystyle
             e_f
         \\[3ex]
L_Z(f) & = & \displaystyle
             \frac{I_{3 f} - \sin^2 \theta_W e_f}{\sin \theta_W \cos \theta_W}
         & & 
R_Z(f) & = & \displaystyle - \tan \theta_W e_f
\end{array}
	\label{SM_ampl}
\end{equation}
where $e_{f}$ is the charge of the fermion and $I_{3f}$ its weak 
isospin. Clearly, the amplitudes for scattering 
off quarks and antiquarks are related by 
\begin{equation}
   A^{\mathrm{SM}}_{a L} (\bar q) = - A^{\mathrm{SM}}_{a R} (q)  \mand
   A^{\mathrm{SM}}_{a R} (\bar q) = - A^{\mathrm{SM}}_{a L} (q)
\end{equation}
A similar relation obtains for $\mu^+$ scattering.
The presence of \rpv\ interactions can introduce either 
$s$-channel (resonance) or $u$-channel diagrams depending on the 
nature of coupling as well as the parton being scattered. 
A brief examination of eqns.\ (\ref{eq:LQD}) shows that the only amplitudes to 
be modified are:
\begin{equation}
\begin{array}{rcl c rcl}
A_{LL}(u_j) & \longrightarrow &  \displaystyle A^{\mathrm{SM}}_{LL}(u) 
                + \frac{\lambda^{\prime 2}_{2 j k} }
                       { \hat s - m^2_{d_{Rk}} + i m_{d_{Rk}} \Gamma_{d_{Rk}}}
         & \qquad & 
A_{LR}(\bar u_j) & \longrightarrow &  \displaystyle A^{\mathrm{SM}}_{LR}(\bar u) 
                + \frac{\lambda^{\prime 2}_{2 j k} }
                       { \hat u - m^2_{d_{Rk}} }
	\\[3ex]
A_{LL}(\bar d_j) & \longrightarrow &  \displaystyle A^{\mathrm{SM}}_{LL}(\bar d) 
                +\frac{\lambda^{\prime 2}_{2 k j} }
                       { \hat s - m^2_{u_{Lk}} + i m_{u_{Lk}} \Gamma_{u_{Lk}}}
         & \qquad &
A_{LR}(d_j) & \longrightarrow & \displaystyle A^{\mathrm{SM}}_{LR}(d) 
				   + \frac{\lambda^{\prime 2}_{2 k j}}
					  { \hat u - m^2_{u_{Lk}} }
\end{array}\; ,
	\label{Rp_ampl}
\end{equation}
where $m_{q_{L,R\:k}}$ is the mass of the exchanged squark.
Given equations (\ref{SM_ampl}) and (\ref{Rp_ampl}), it is easy to 
calculate the pure standard-model contribution (the background) as well as the 
pure-\rpv\ and interference terms, which together comprise the signal.  
The expressions in eqn.\ (\ref{Rp_ampl}) correspond to the case when 
only one of the \rpv\ couplings is nonzero.\footnote{For a more general 
case, the right-hand sides are replaced by a sum over the relevant 
couplings (and corresponding squark masses).  We do not consider such 
cases in this work.}

Before we attempt to disentangle the signal from the background, we 
must consider how detector characteristics may limit
measurements of this specific process. 
Excluding a cone of half-angle $6^{\circ}$ around the muon beam 
appears necessary for 
muon identification~\cite{HSmumu,SRmumu}. Similarly, accurate 
measurements involving jets that lie very close to the beam pipe 
seems unlikely. Thus, we shall require that 
\begin{equation}
10^{\circ} < \theta_{J} < 170^{\circ}   \ ,
	\label{j_cuts}
\end{equation}
where $\theta_J$ is the polar angle of the struck parton in the
final state (identifiable with the thrust axis of the
final-state monojet), and
\begin{equation}
15^{\circ} < \theta_{\mu} < 165^{\circ} \ .
	 \label{mu_cuts}	
\end{equation}
Such {\em acceptance cuts} significantly reduce the standard-model 
background, for it is primarily peaked in the forward direction.  On 
the other hand, the \rpv\ signal due to an isotropically decaying 
squark resonance is much less peaked.  Since a heavy squark is 
preferentially produced with a small momentum, the decay-product 
($\mu$ and a jet) distributions are nearly isotropic.

\begin{figure}[htb]
         \centerline{\BoxedEPSF{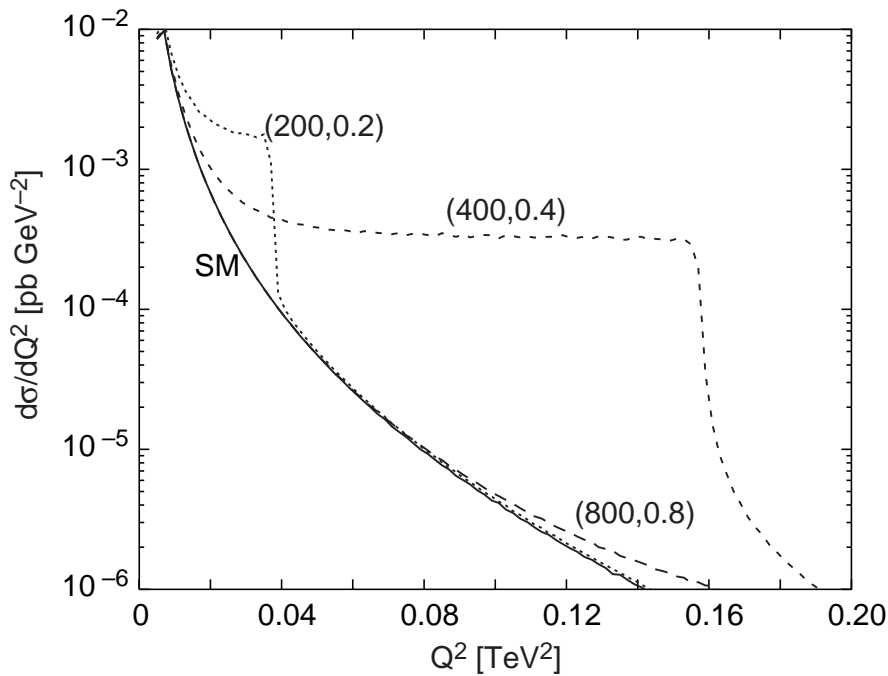 scaled 600}} 
\vspace*{10pt}
\caption{The $Q^2$-distribution for the neutral current
	process $\mu^+ + p \to \mu^+ + X$ at the
	$(200 \gev \times 1 \tev)$ machine. The solid line
	represents the standard-model expectations, while the other
	curves are for the displayed values of
	($m_{\tilde u_{Lj}}, \lambda^{\prime}_{2j1}$). The only cuts
	are those of eqns. (\protect\ref{j_cuts}) and (\protect\ref{mu_cuts}).
	}\
\label{fig:nc_qsq}
\end{figure}

According to the amplitudes (\ref{Rp_ampl}), the only possible 
resonances involving a valence quark are ($i$) $\mu^+ + d \to \tilde 
u_{Lj}$ when $\lambda^{\prime}_{2 j 1} $ is nonzero, and ($ii$) $\mu^- 
+ u \to \tilde d_{Rk}$ when $\lambda^{\prime}_{2 1 k} $ is nonzero.  
For definiteness, we restrict ourselves here to the first alternative.  
The sensitivity to the other channel is qualitatively similar (in 
fact, even somewhat greater).  We show in Figure~\ref{fig:nc_qsq} the 
$Q^{2}$ distribution for the production of $\tilde u_{Lj}$ at the 
$200\gev \times 1\tev$ machine, for several choices of parameters.

\begin{figure}[htb]
% \vspace*{-1.5cm}
% \hspace*{-2.3cm}
% \begin{center}
         \centerline{\BoxedEPSF{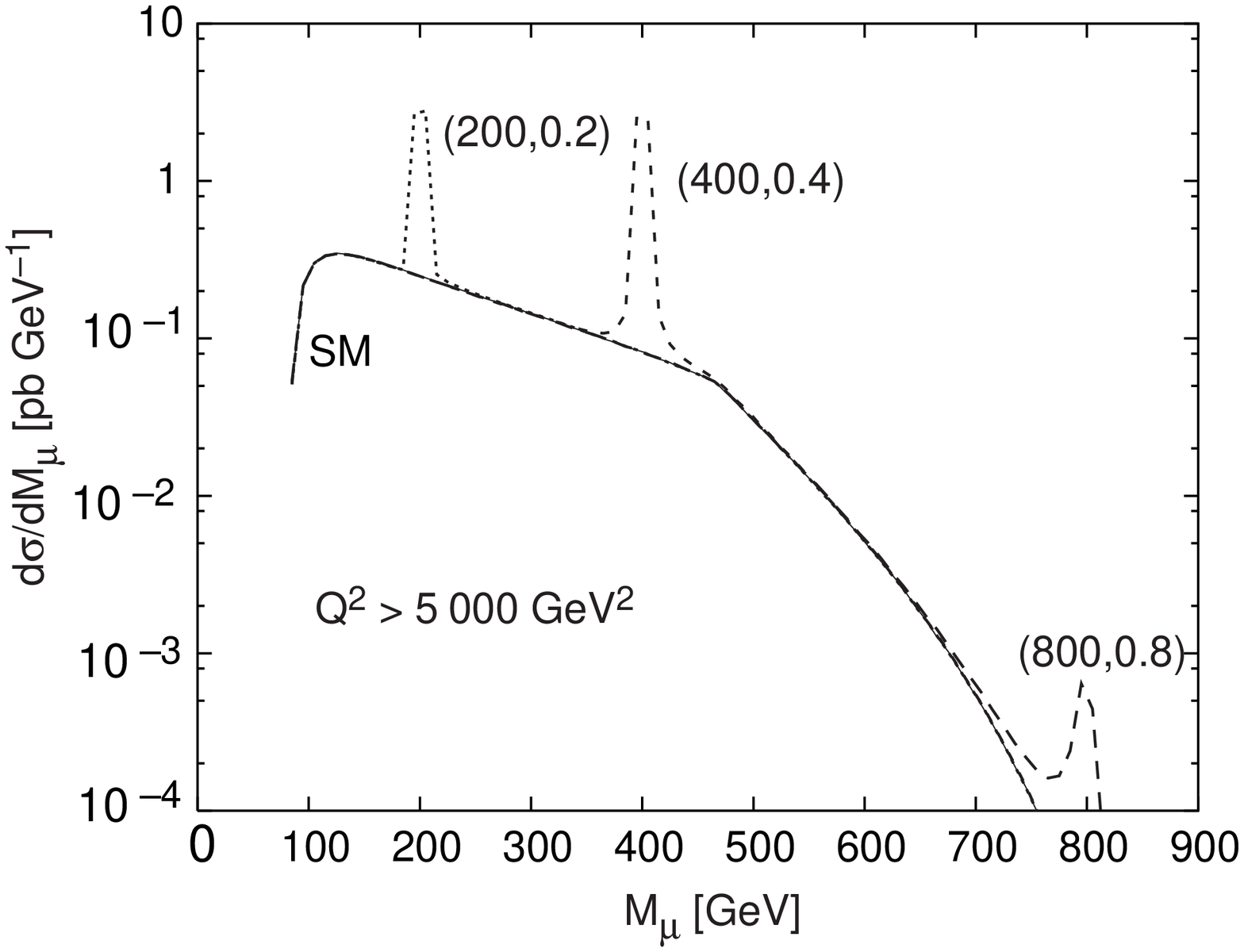  scaled 
         300}\quad\BoxedEPSF{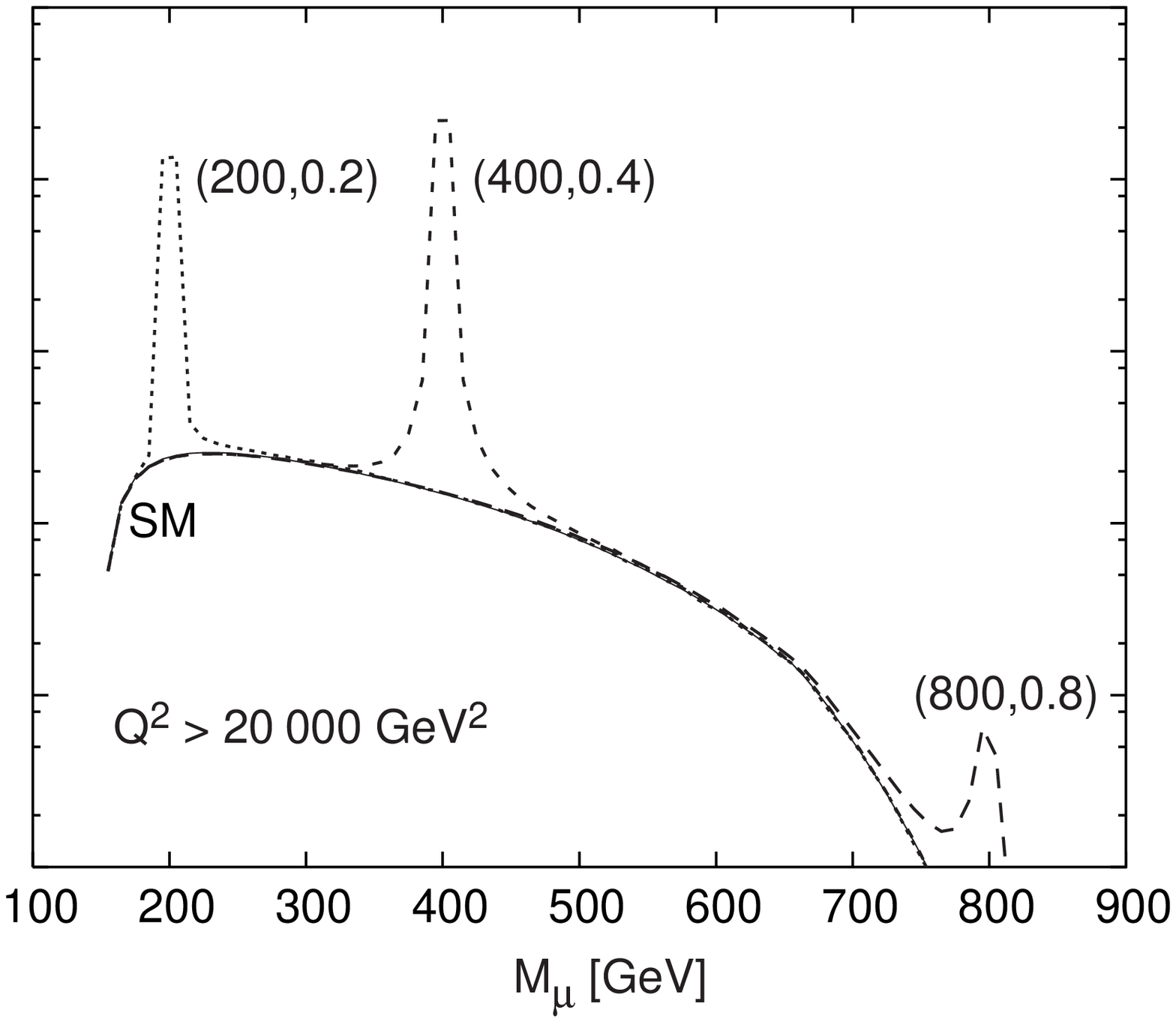  scaled 300} 
         \quad\BoxedEPSF{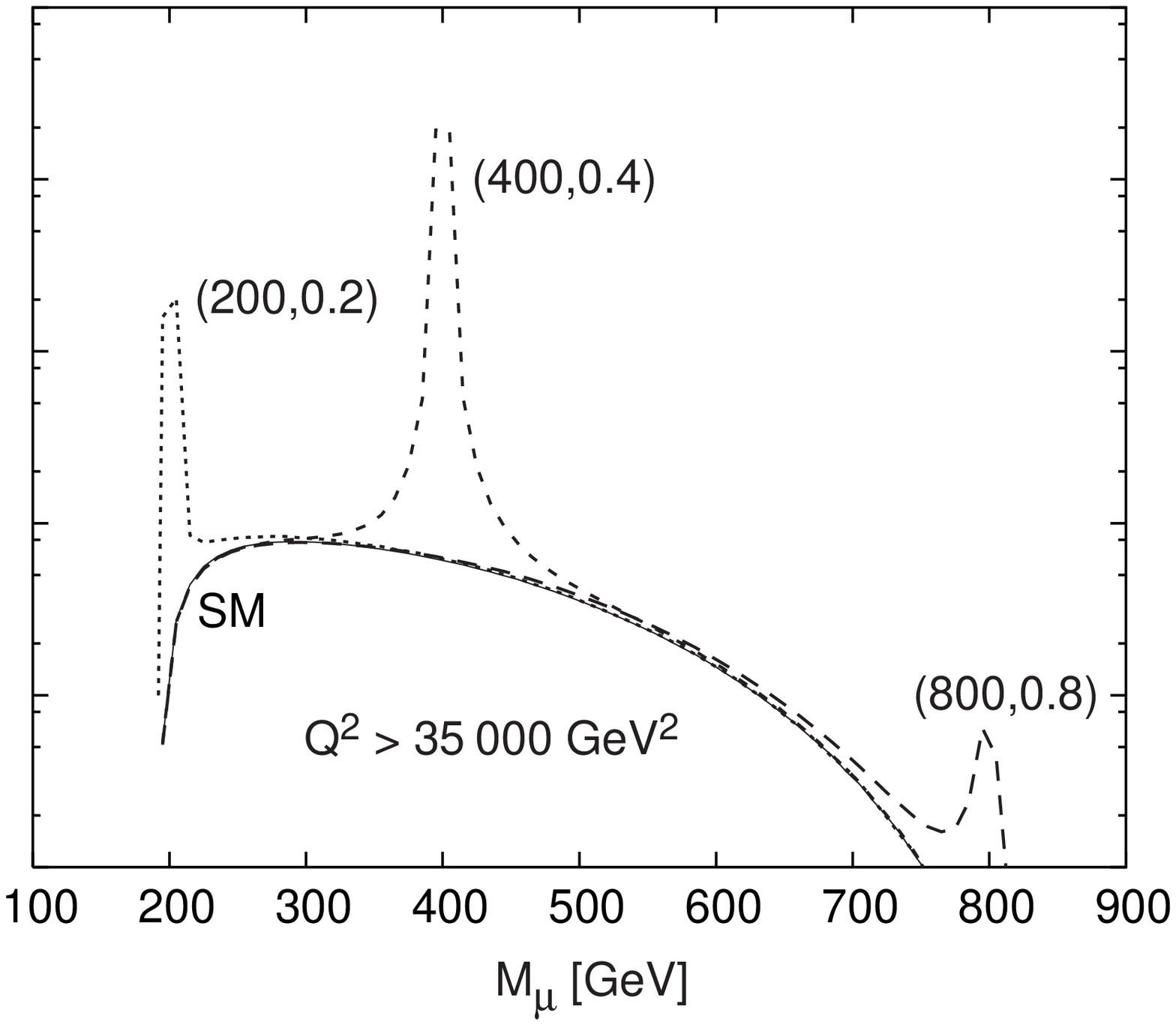 scaled 300}}    
\vspace*{10pt}
\caption{The invariant mass distribution for the neutral current 
process $\mu^+ + p \to \mu^+ + X$ at the $(200 \gev \times 1 \tev)$ 
machine.  The solid line represents the standard-model expectations, 
while the other curves are for the displayed values of $(m_{\tilde 
u_{Lj}}, \lambda^{\prime}_{2j1})$.  In addition to the cuts of 
eqns. (\protect\ref{j_cuts}) and (\protect\ref{mu_cuts}), we impose a 
cut on $Q^2$.  }
\label{fig:nc_mass}
\end{figure}

The standard-model contribution peaks at low 
scattering angles, or, in other words, at low $Q^2$ values.  In 
contrast, the signal events typically populate a much larger $Q^2$ 
range.  A look at Figure~\ref{fig:nc_qsq} thus suggests that harder 
$Q^2$ cuts would tend to enhance the signal-to-noise ratio.  
The presence of possible squark resonances suggests that distributions 
in invariant mass $M_\mu$ of the final (muon + monojet) state would be 
sensitive to the new physics effects.  In Figure~\ref{fig:nc_mass}, we 
present this distribution for three different cases obtained by 
imposing different cuts on the minimum $Q^2$.  As expected, the 
resonances stand out sharply, and are little affected by the $Q^2$ 
cut, while the background is strongly suppressed when we demand a 
higher $Q^2$ threshold.

\begin{figure}[htb]
         \centerline{\BoxedEPSF{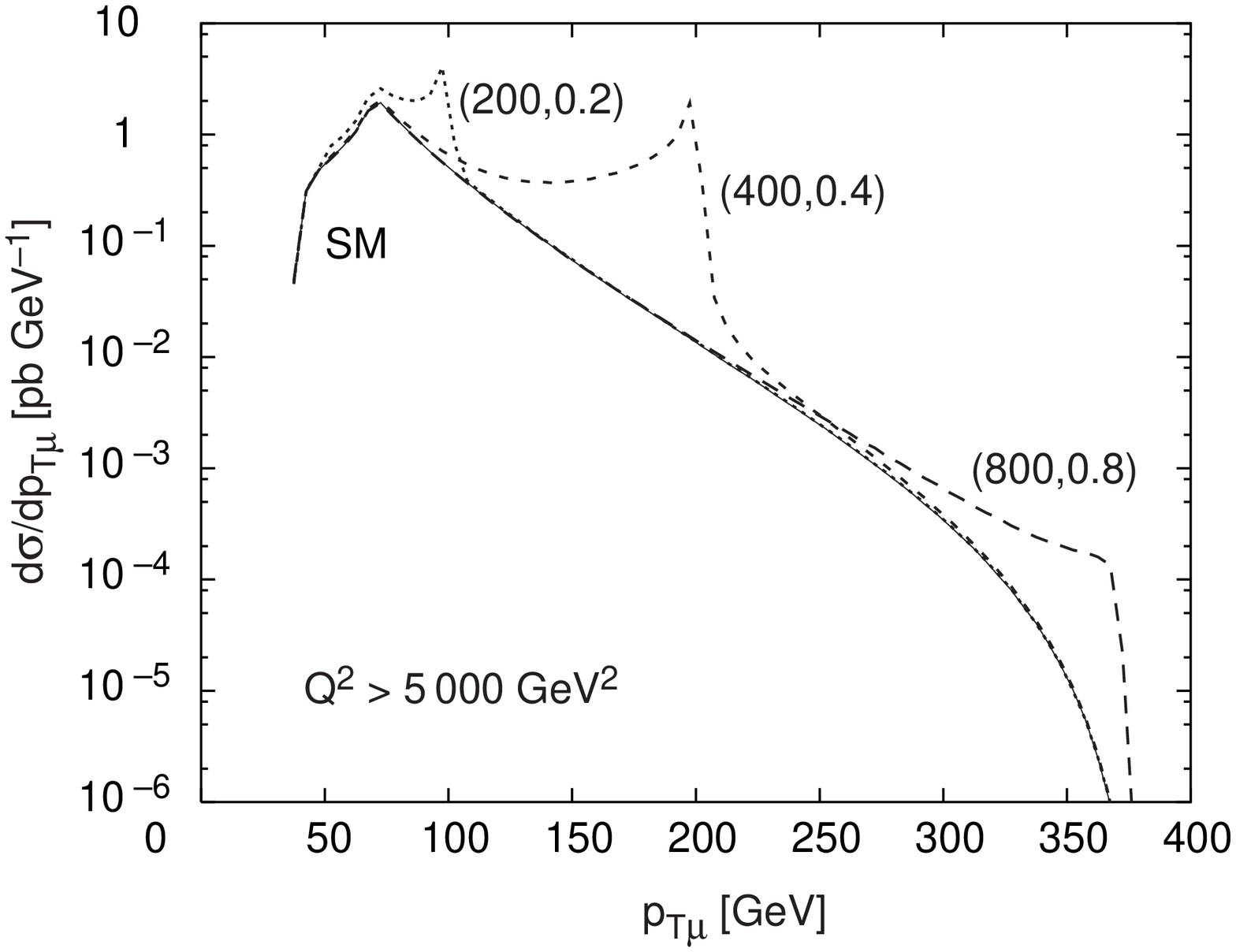  scaled 
         300}\quad\BoxedEPSF{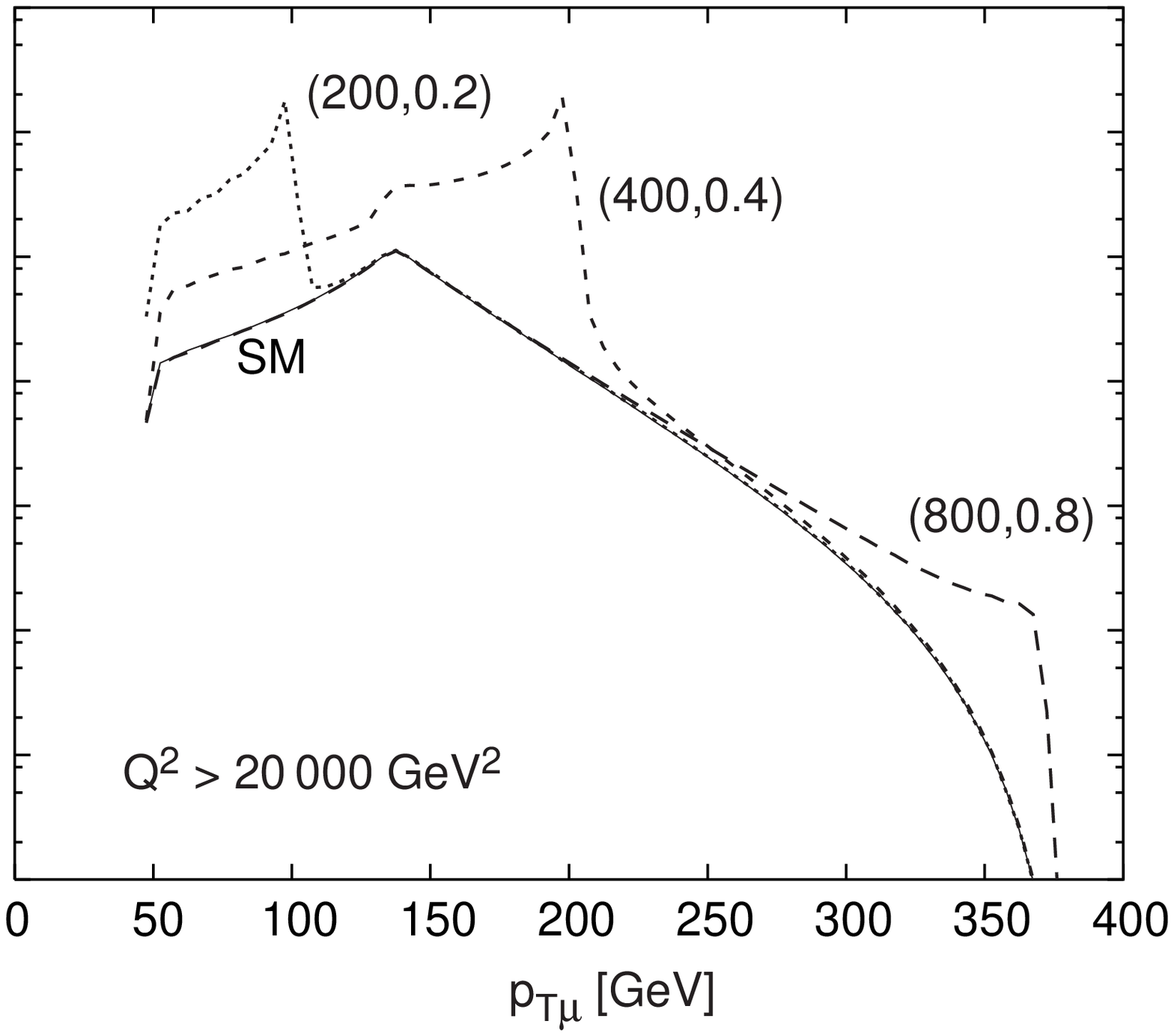  scaled 300} 
         \quad\BoxedEPSF{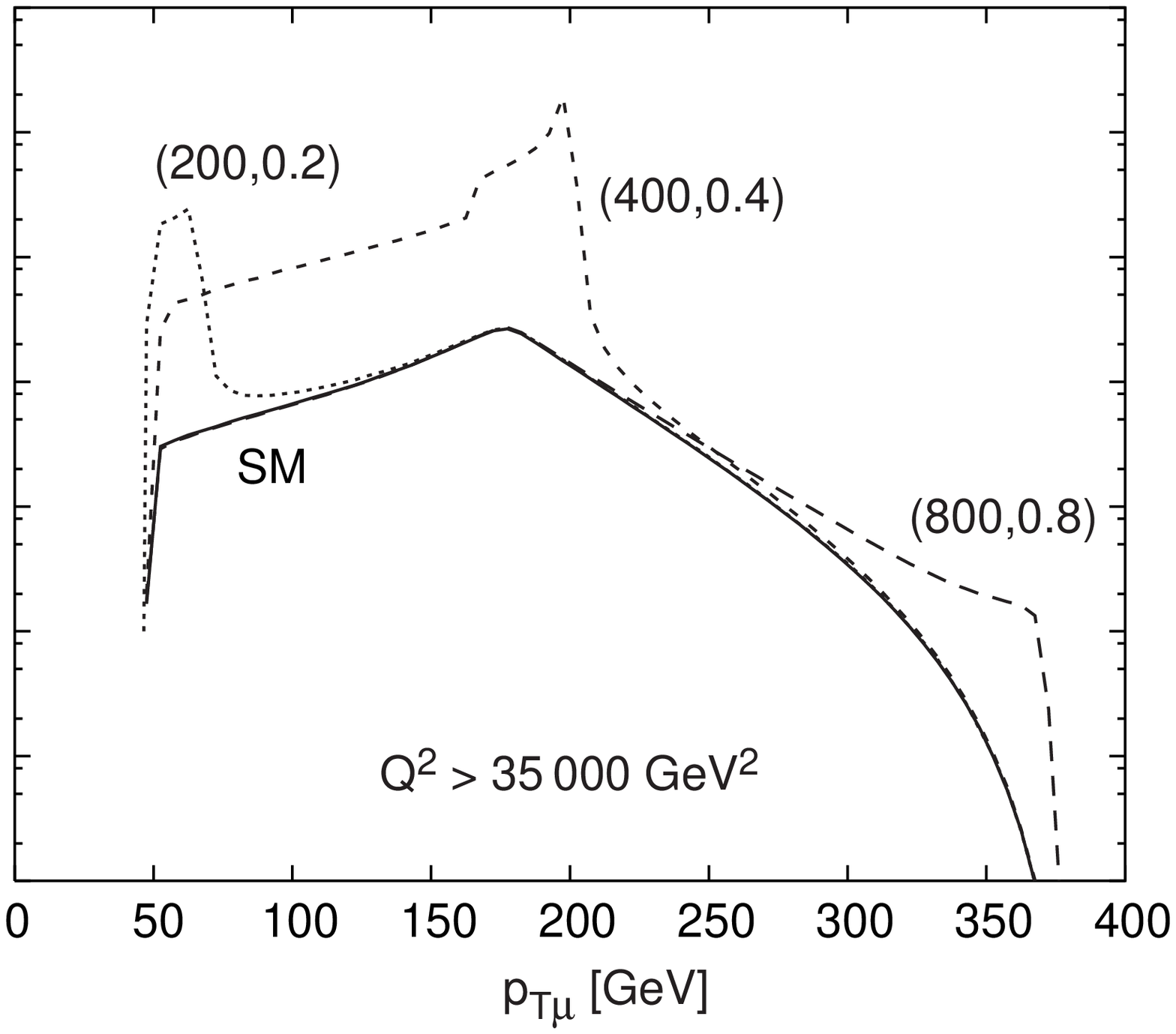 scaled 300}}    
\vspace*{10pt}
\caption{The transverse momentum distribution for the neutral current 
	process $\mu^+ + p \to \mu^+ + X$ at the 
	$(200 \gev \times 1 \tev)$ machine. The solid line
	represents the standard-model expectations, while the other 
	curves are for the displayed values of 
	$(m_{\tilde u_{Lj}}, \lambda^{\prime}_{2j1})$. In addition to the cuts of 
eqns.(\protect\ref{j_cuts}) and (\protect\ref{mu_cuts}), we impose a 
cut on $Q^2$. 
        } 
\label{fig:nc_pT}
\end{figure}

A similar, though not so sharp, excess can also be seen in the $p_T$ 
distribution shown in Figure~\ref{fig:nc_pT}.  The sharp fall-off of 
the excess is a manifestation of the well-known Jacobian peak.  The 
price of overzealous $Q^{2}$ cuts is to discard a large fraction of 
the signal in the interest of suppressing the background.  For 
example, a cut of $Q^2 > 35,000 \gev^2$ accentuates the signal due to 
a $800\hbox{-GeV}\!/\!c^{2}$ squark, but eliminates most of the signal 
due to a $200\hbox{-GeV}\!/\!c^{2}$ squark (see 
Figure~\ref{fig:nc_mass}$c$ or Figure~\ref{fig:nc_pT}$c$).  The 
optimum value of the $Q^2$ cut is thus a sensitive function of the 
squark mass.

Rather than design a mass-specific cut, we opt to use the difference 
in the distributions in a slightly different, but more efficient way. 
We divide the $Q^2$--$M_\mu$ plane into equal-sized bins, and 
compute the number of signal ($N_{n}^{{\mathrm{SM}} + \not{R}}$)
and background ($N_{n}^{\mathrm{SM}}$)   events in each bin $n$
for an accumulated luminosity of $1\fb^{-1}$. We then define a $\chi^2$
test of discrimination 
\begin{equation}
   \chi^{2} = \sum_{n} \frac{\left(N_{n}^{{\mathrm{SM}}+ \not{R}} 
                               - N_{n}^{{\mathrm{SM}}}\right)^{2}}
	{N_{n}^{\mathrm{SM}} + (\epsilon N_{n}^{\mathrm{SM}})^2 }
	\label{eq:chisq}
\end{equation}
where $\epsilon$ is a measure of the systematic error, accruing mainly 
from the uncertainties in luminosity measurement and parton densities.  
To be specific, we use a uniform grid of ($4000\gev^2, 40\gevcc$) and 
perform the sum over all the bins for which $N_n^{\mathrm SM} \ge 1$. 
For the systematic error, we choose $\epsilon = 5 \%$.  
As it turns out, the final results are not too sensitive to the choice of 
$\epsilon$.

\begin{figure}[htb]
         \centerline{\BoxedEPSF{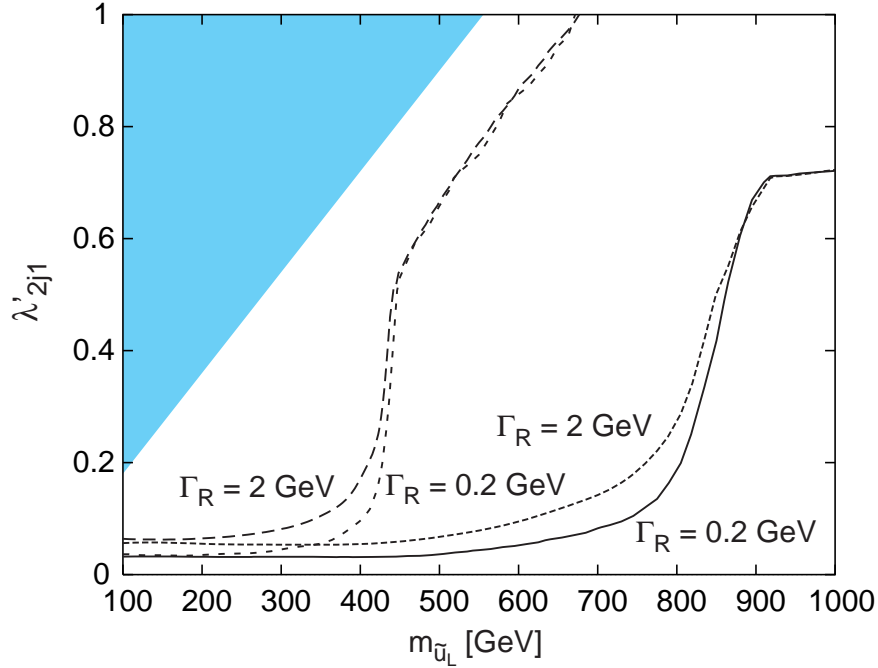  scaled 600}}    
\vspace*{10pt}
\caption{Exclusion contours that may be obtained from neutral-current 
processes at a $\mu^+ p$ collider with an accumulated luminosity of $1 
\fb^{-1}$.  The part of the parameter space above the curves may be 
ruled out at the 95\% C.L.  The set on the left corresponds to the 
$(50 \gev \times 1\tev)$ mode while that on the right corresponds to 
a $(200\gev \times 1\tev)$ machine.  For each case, the dependence 
on the $R$-conserving width is also shown.  The shaded region 
corresponds to the area ruled out by low-energy experiments.  The 
corresponding right-handed squark is assumed to have a mass of $2\tevcc$.  }
\label{fig:nc_excl}
\end{figure}

In Figure~\ref{fig:nc_excl}, we illustrate the reach of such an 
experiment in the $m_{\tilde u_{Lj}}$-$\lambda^{\prime}_{2j1}$ plane.  
The region of the parameter space {\em above} the individual curves 
can be ruled out at the 95\% C.L.  Alternatively, for a given value of 
one of the two parameters, the corresponding projection onto the other 
axis gives the 98.6\% C.L.\ limit on the other parameter.  To obtain an 
understanding of the curves, it is instructive to consider only the 
resonant contribution, which goes as
\begin{equation} \displaystyle
    \sigma (\mu^+ + d \to \tilde u_L) 
	   = \frac{\pi \lambda_{2 j 1}^{\prime\:2} }{4 \pi s_{\mu p}}  \;
	      f_d\left(\frac{m^2_{\tilde u_L} }{s_{\mu p}}, s_{\mu p}\right) \ ,
\end{equation}
where $f_d$ is the density of the $d$-quark at the corresponding value 
of the Bjorken variable $x = m^2_{\tilde u_L} / s_{\mu p}$ and 
virtuality $s_{\mu p}$.  The exclusion curve obtained from this piece 
alone would read
\begin{equation} 
 \lambda_{2 j 1}^{\prime\:2}\: {\cal B}(\tilde u_L \to \mu^+ + d) \: 
	f_d\left(\frac{m^2_{\tilde u_L} }{s_{\mu p}}, s_{\mu p}\right) 
 = {\mathrm constant,}
\end{equation} 
where ${\cal B}$ is the branching fraction for squark decay into the 
observed channel.  Fixing the parton distributions, we can determine 
the analytic dependence of the exclusion curves on the parameters.  
For example, the effect of the $R$-conserving width $\Gamma_R$ can 
be understood from this relation.  Since the \rpv\ width goes as 
$\lambda^{\prime\:2}$, for small values of ${\cal B}$, the exclusion 
curve in $\lambda^{\prime}$ goes as $\Gamma_R^{1/4}$.  With an 
increase in $m_{\tilde u_L}$, which increases the bound on 
$\lambda^{\prime}$, the \rpv\ width increases and becomes comparable 
to or even dominates over $\Gamma_R$.  This, for example, leads to the 
coalescing of the curves for $\Gamma_R = 0.2\gev$ and $\Gamma_R = 
2\gev$ (see Figure~\ref{fig:nc_excl}).

The $200\gev \times 1\tev$ exclusion plot levels off at high values 
of $m_{\tilde u_L}$.  For very massive 
squarks,  resonance formation is not possible kinematically.  Rather, the major 
effect due to the squark arises in the form of $s$- and $u$-channel 
exchanges that lead naturally to excesses in large $Q^2$, but moderate 
$M_\mu$, regions of the phase space.  Consequently, the dependence on 
the squark mass is less pronounced.  In this regime, the contributions 
we have neglected from a right-handed squark (assumed to have a mass 
of $2\tevcc$) need to be taken into account, especially if it couples 
to a $u$-quark ($\lambda^{\prime}_{211}$).

\subsection{Charged-current interactions}\label{subsec:CCsigs}
The charged-current reaction $\mu^{-}p \to \nu_{\mu}+ X$ is mediated 
by $t$-channel $W$-boson exchange and by $\tilde{d}_{R}^{k}$ 
excitation through the \rpv\ coupling $\lambda^{\prime}_{21k}$.  The 
inclusive cross section is given by
\begin{equation}
\sigma(\mu^-p \to \nu_\mu X) = \sum_{q}
\int_0^1 {\rm d} x \: f_{q}(x, Q^2) \: 
         \hat{\sigma}(\mu^-q \to \nu_\mu q^{\prime}) \ ,
\end{equation}
where $\hat{\sigma}(\mu^-q \to \nu_\mu q^{\prime})$ is the elementary 
cross section for the $\mu^-$ to scatter off quark $q$ with momentum 
fraction $x$.  At the parton level, the only allowed charged current 
processes are
\begin{equation}
	\mu^- + u_i \to \nu_\mu + d_i \mand
	\mu^- + \bar d_i \to \nu_\mu + \bar u_i
\end{equation}
and the conjugate processes for a $\mu^+$ beam.  We neglect processes 
involving the top quark and will also neglect 
Cabibbo-Kobayashi-Maskawa mixing.  Both approximations are excellent 
for the present purpose.  The cross sections are much simpler than 
those for the neutral-current process and are given by ($\hat{t} \equiv - 
Q^2$)
\begin{equation} 
\begin{array}{rcl}
\displaystyle \frac{{\rm d} \hat{\sigma}}{{\rm d} \hat t} 
	(\mu^- + q \to \nu_\mu + \bar q) & = &
	 \displaystyle   \frac{1}{16 \pi \hat s^2} \;
		\left| \frac{e^2}{s_W^2 \: (\hat t - m_W^2)}  
			+ A_{\not R}
		\right|^2
	\\[3ex]
A_{\not R} (u_i) & = & \displaystyle
		\frac{\lambda_{2 i k}^{\prime 2} }
		     {\hat s - m_{\tilde d_{k R}}^2 
			+ i \Gamma_{\tilde d_{k R}} m_{\tilde d_{k R}} }
	\\[3ex]
A_{\not R} (\bar d_i) & = & \displaystyle
		\frac{\lambda_{2 i k}^{\prime 2} }
		     {\hat u - m_{\tilde d_{k R}}^2 }
\end{array}
	\label{cc:cross-sec}
\end{equation}
Only a $\mu^-$ beam excites a resonance in charged-current 
interactions with a valence quark.  The resonance contributes equally 
to the charged- and neutral-current processes.  Furthermore, since the 
$u$-quark density in the proton is roughly twice that of the 
$d$-quark, the effect of the smaller branching fraction is compensated 
to a great extent.  Thus, we could have used this operation mode for 
the neutral-current process as well, with the difference that we would 
now explore a different coupling.

\begin{figure}[htb]
         \centerline{\BoxedEPSF{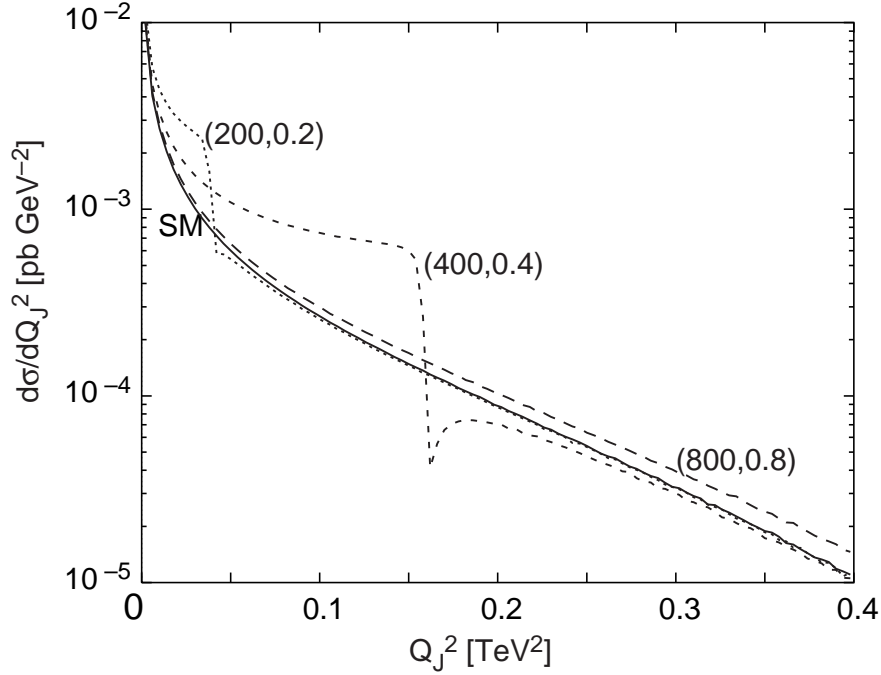  scaled 600}}    
\vspace*{10pt}
\caption{The $Q^2$-distribution for the charged current process $\mu^- 
+ p \to \nu_\mu + X$ at the $(200 \gev \times 1 \tev)$ machine.  The 
solid line represents the standard-model expectations, while the 
other curves are for the displayed values of $(m_{\tilde d_{Rk}}, 
\lambda^{\prime}_{21k})$.  The only cut is that of 
eqn.\ (\protect\ref{j_cuts}).  }\
\label{fig:cc_qsq}
\end{figure}

Because the charged-current process has no photon-exchange 
contribution, the forward peak of the standard-model background is 
significantly reduced (see Figure~\ref{fig:cc_qsq}) compared to the 
neutral-current case.  This implies that the deviation due to the 
presence of an \rpv\ coupling should be visible even at smaller $Q^2$ 
values.  However, since the neutrino cannot be seen directly, we have 
only the relatively crude measure of $Q_J^2$ at our disposal, so the 
$Q^2$ plateau resulting from the resonance is degraded compared with 
the neutral-current case.

\begin{figure}[htb]
         \centerline{\BoxedEPSF{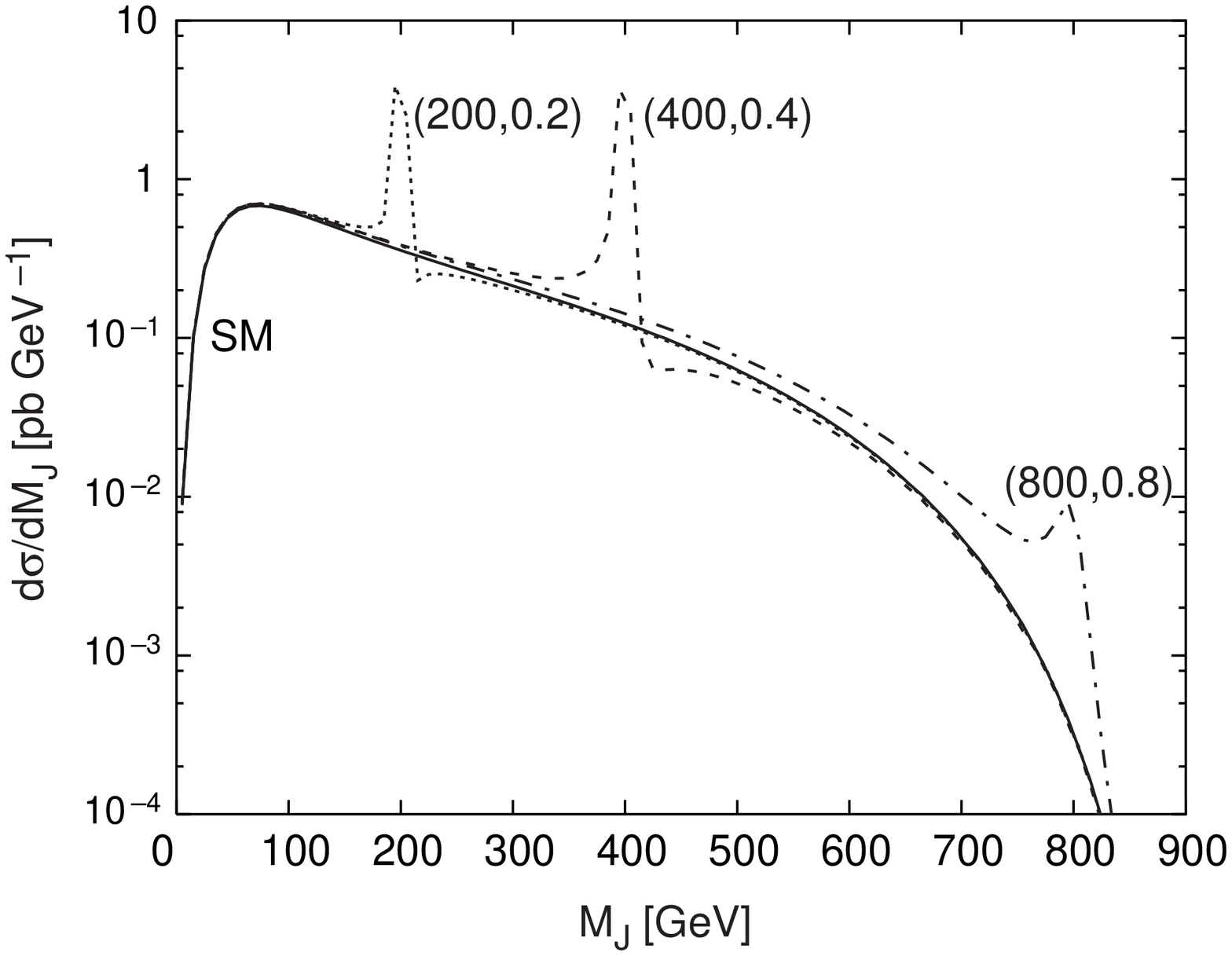  scaled 
         300}\quad\BoxedEPSF{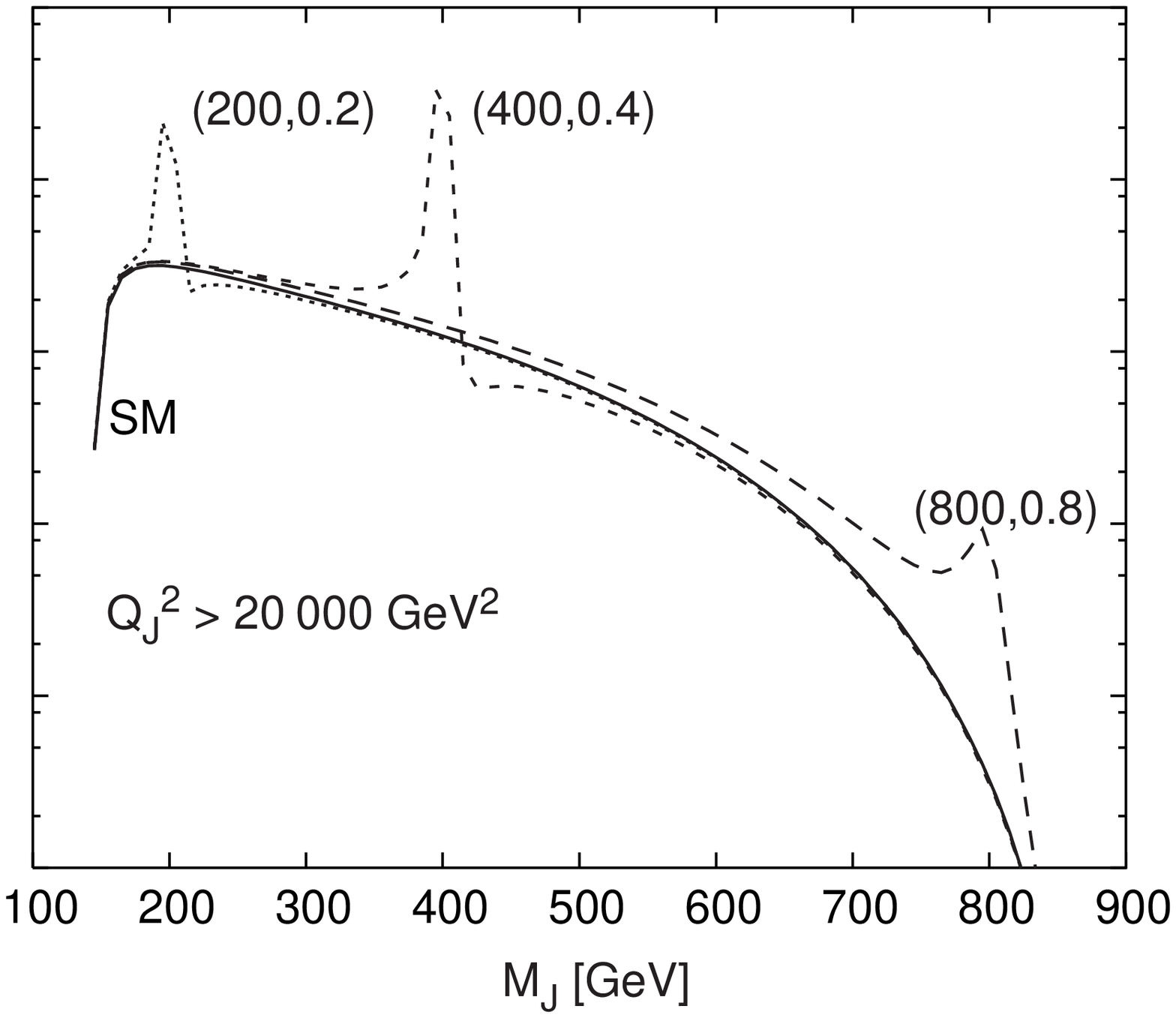  scaled 300}
	 \quad\BoxedEPSF{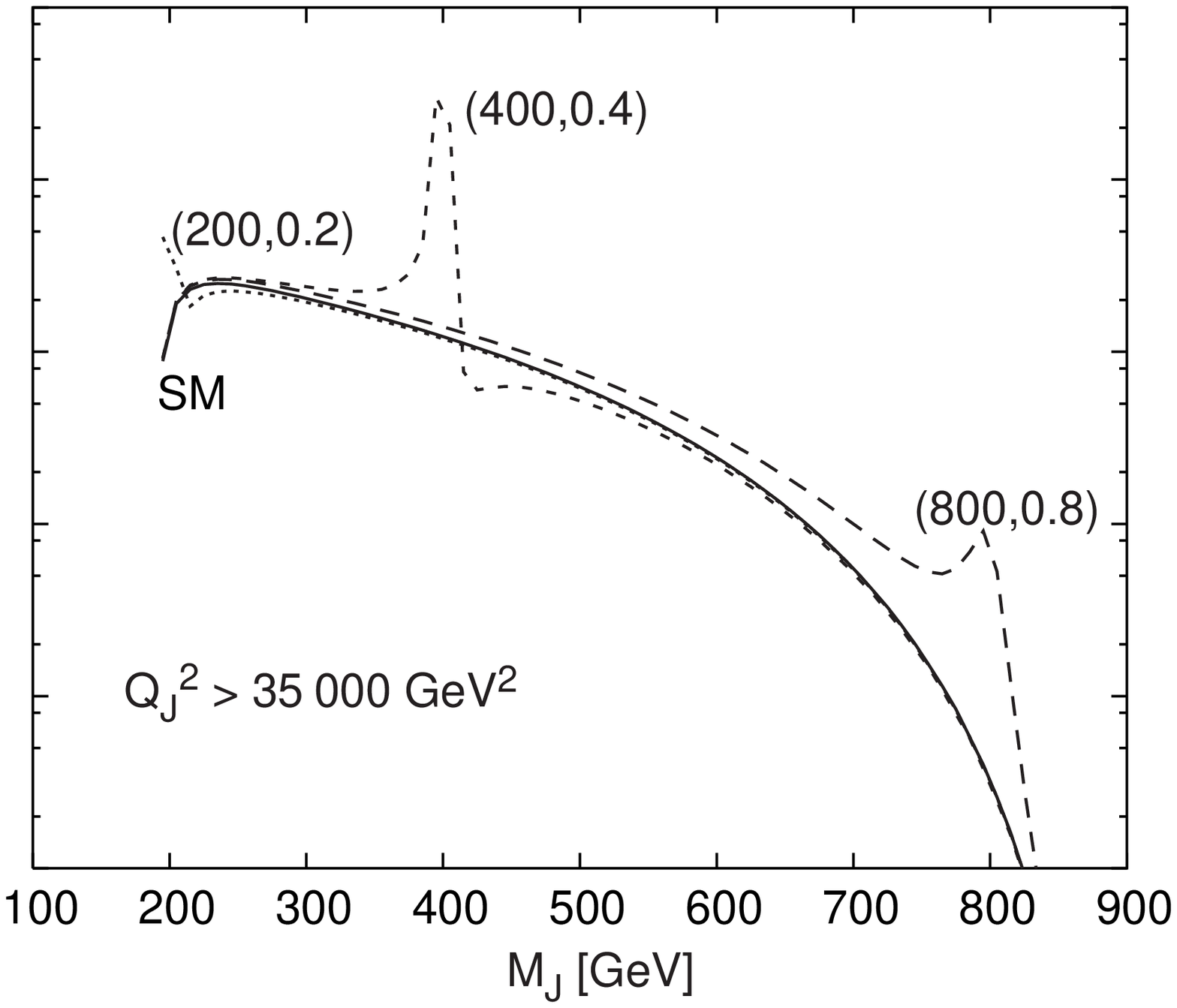  scaled 300}}    
\vspace*{10pt}
\caption{The invariant mass distribution for the charged-current 
process $\mu^- + p \to \nu_\mu + X$ at the $(200\gev \times 1\tev)$ 
machine.  The solid line represents the standard-model expectations, 
while the other curves are for the displayed values of $(m_{\tilde 
d_{Rk}}, \lambda^{\prime}_{21k})$.  In addition to the cut of 
eqn.\ (\protect\ref{j_cuts}), we impose a cut on $Q_J^2$.  }
\label{fig:cc_mass}
\end{figure}

\begin{figure}[htb]
         \centerline{\BoxedEPSF{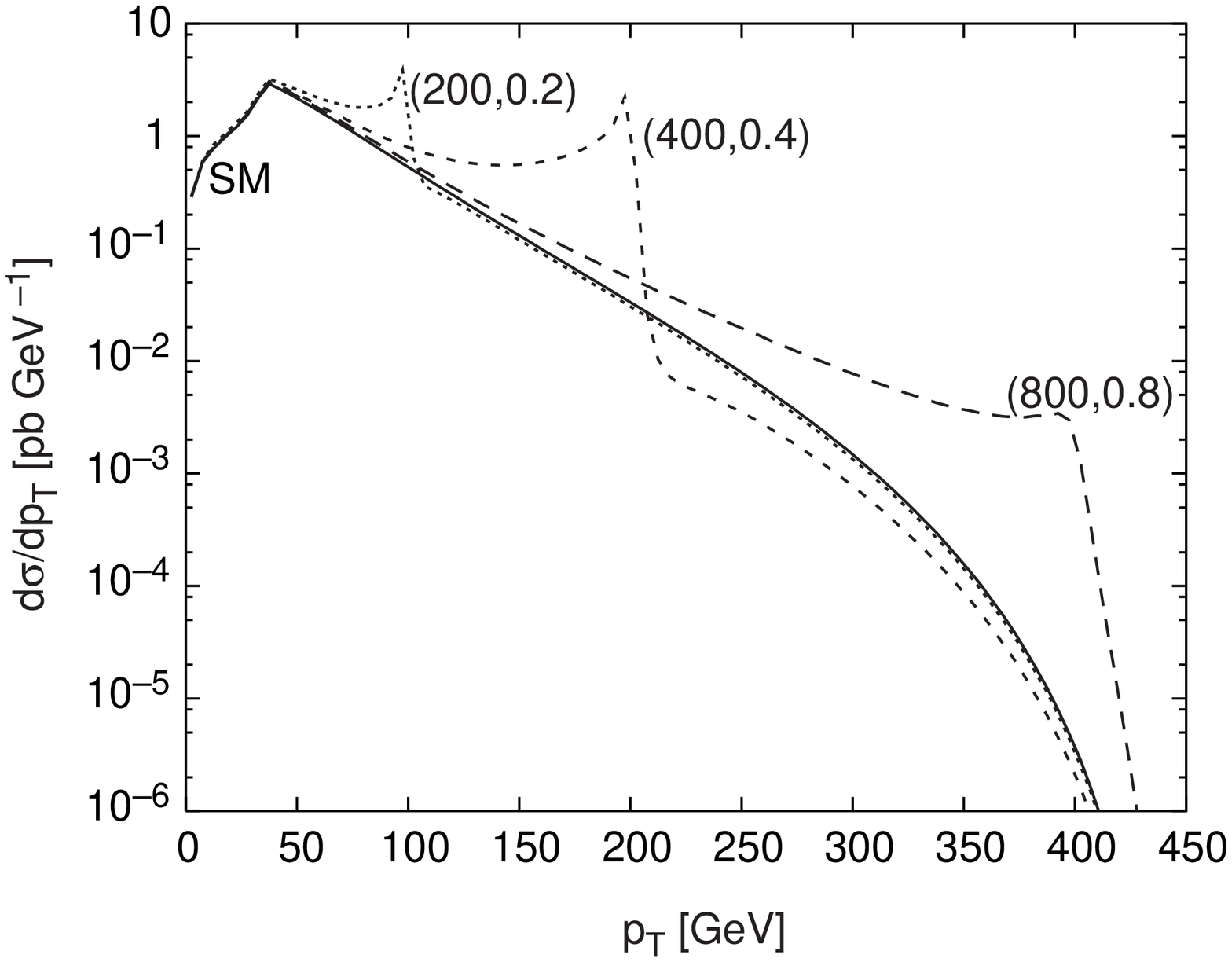  scaled 
         300}\quad\BoxedEPSF{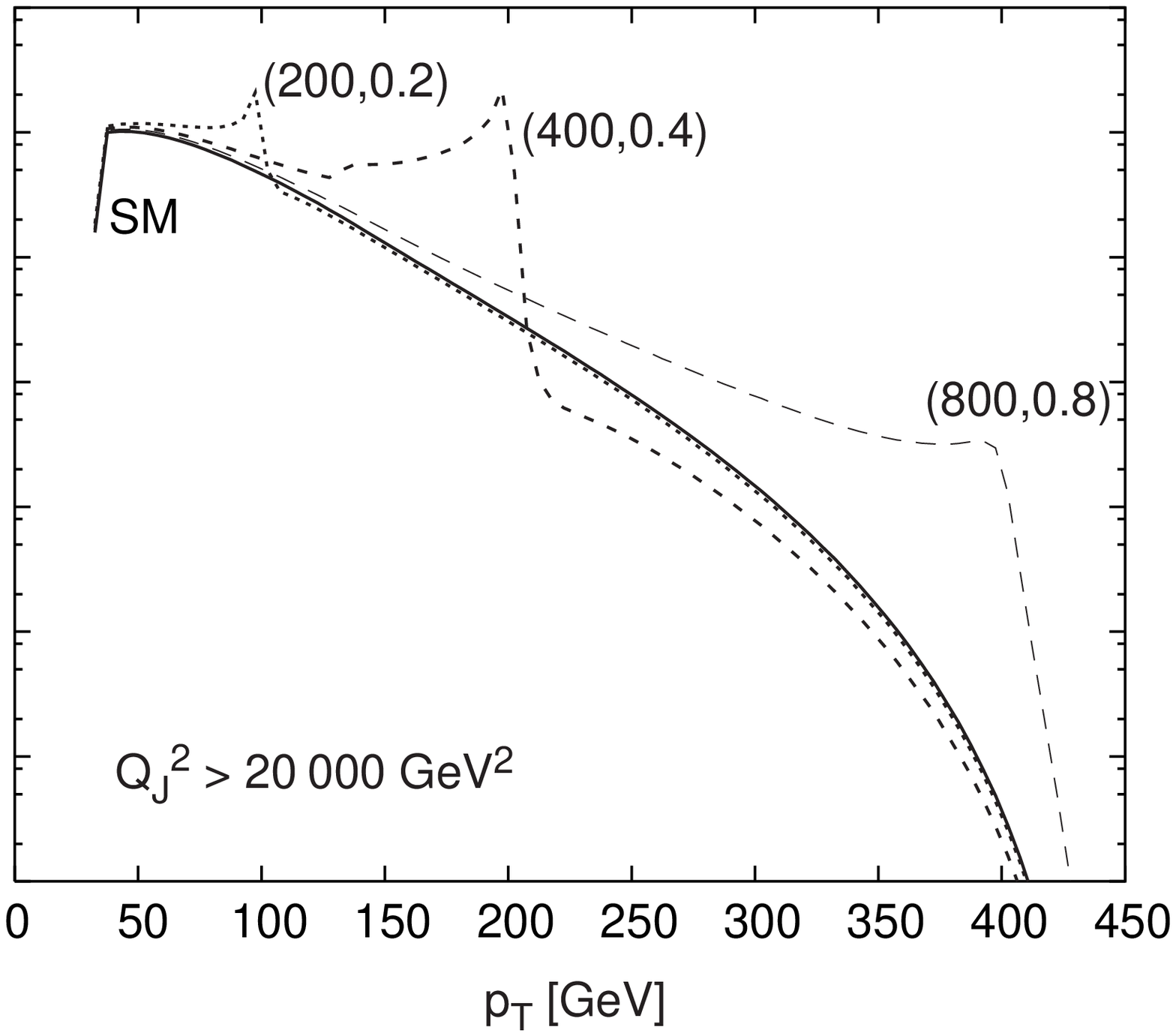  scaled 300}
	 \quad\BoxedEPSF{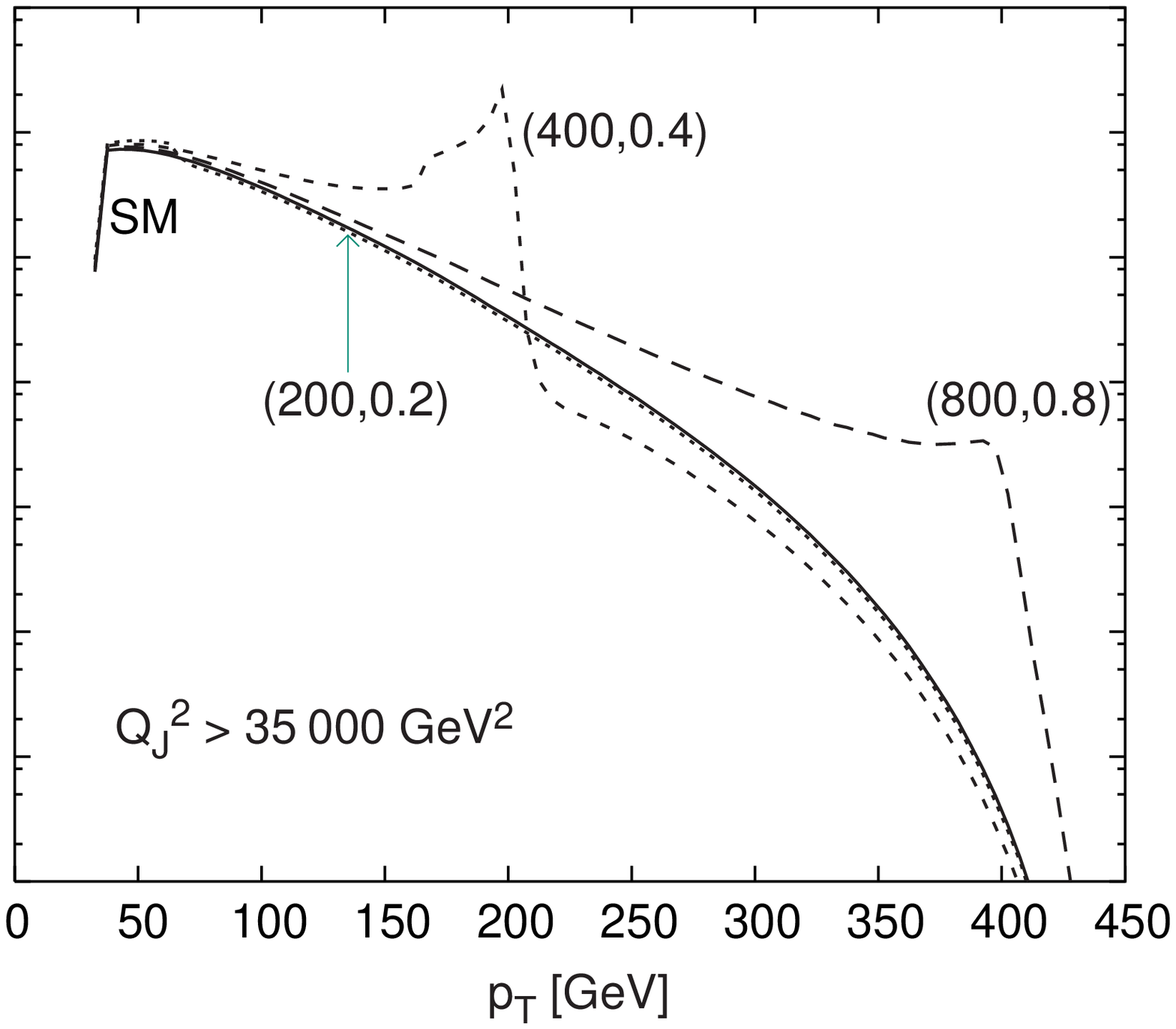  scaled 300}}    
\vspace*{10pt}
\caption{The jet transverse momentum distribution for the 
charged-current process $\mu^- + p \to \nu_\mu + X$ at the $(200 \gev 
\times 1 \tev)$ machine.  The solid line represents the standard-model 
expectations, while the other curves are for the displayed values 
of $(m_{\tilde d_{Rk}}, \lambda^{\prime}_{21k})$.  In addition to the 
cut of eqn.\ (\protect\ref{j_cuts}), we impose a cut on $Q_{J}^2$.}
 \label{fig:cc_pT}
\end{figure}

The distributions in the resonance mass $M_{J}$ reconstructed from jet 
variables (Figure~\ref{fig:cc_mass}), and in the jet transverse 
momentum $p_{TJ}$ (Figure~\ref{fig:cc_pT}) are qualitatively similar 
to those for the neutral-current case.  In our examples, the \rpv\ 
term interferes constructively with the standard-model background 
below resonance, and destructively above.  

\begin{figure}[htb]
         \centerline{\BoxedEPSF{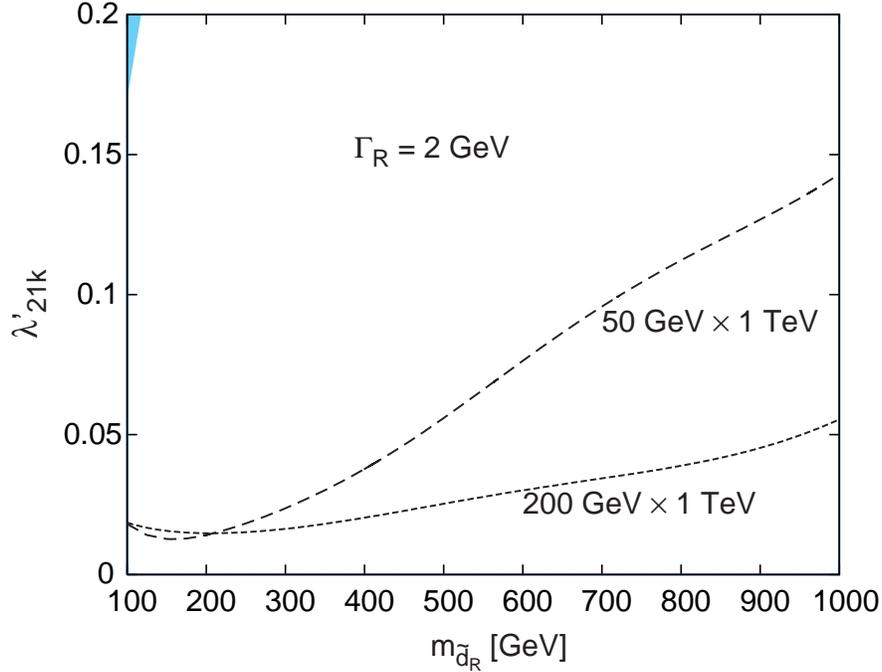  scaled 
         600}}    
\vspace*{10pt}
\caption{Exclusion contours that may be obtained from the 
charged-current process at a $\mu^- p$ collider with an accumulated 
luminosity of $1 \fb^{-1}$.  The part of the parameter space above the 
curves may be ruled out at 95\% C.L.  The tiny shaded region in the 
northwest corner corresponds to the area ruled out by low-energy 
experiments.  The corresponding left-handed squarks are assumed to 
have a mass of $2\tevcc$.  }
\label{fig:cc_excl}
\end{figure}

To determine the reach of this experiment, we use a similar binning as 
before, only now in the ($Q_J^2, M_J$) plane.  The resultant contours 
are presented in Figure~\ref{fig:cc_excl}.  The generic features are 
quite similar to those of Figure~\ref{fig:nc_excl}, but the 
sensitivity is greatly increased, notwithstanding the smaller number 
of kinematic observables.  The reasons for the greater sensitivity are 
easy to see:
\begin{itemize}
   \item[$\bullet$]
	The photon-exchange contribution, the major background in 
	the neutral-current mode, is absent here. 
   \item[$\bullet$]The dominant standard-model charged-current subprocess 
   in $\mu^- p$ scattering involves the up-quark.  With our choice of 
   couplings, the dominant \rpv\ amplitude also involves the same quark.  
   Thus, the interference term is maximized.
   \item[$\bullet$]The apparent advantage of the charged-current channel 
   is due, in part, to our choice of the initial state.  Had we 
   considered neutral-current processes in $\mu^- p$ scattering instead, 
   the dominant \rpv\ process would have been resonant production of 
   $\tilde{d}_{R}^{k}$.  The smaller branching 
   fraction of the $\tilde{d}_{R}^{k}$ into $\mu^-$ is more than 
   compensated by the larger flux of the $u$-quark.  Moreover, the 
   standard-model amplitude for $\mu^- u $ scattering being larger than 
   that for $\mu^+ d $ scattering, the relative importance of the 
   interference term is larger, especially for off-resonance 
   contributions.  Hence the $\mu^- p$ neutral-current exclusion curves 
   would be slightly stronger than those displayed in 
   Figure~\ref{fig:nc_excl} for $\mu^+ p$.
\end{itemize}

\section{Summary and outlook}\label{sec:conc}
A high-energy, high-luminosity muon-proton collider would offer very 
interesting new possibilities to search for signals of new physics at 
high $Q^{2}$.  To develop one example in some detail, we have 
examined the sensitivity of a 50- or 200-GeV ($E_{\mu}$) $\times$ 1-TeV 
($E_{p}$) collider to \rpar--violating couplings.  We have considered 
situations in which only $LQD$ \rpv\ couplings are nonzero. Prominent signals 
would arise from the resonant formation of a squark that decays into 
a hadron jet plus a muon or a neutrino.  For a  squark 
with mass $\alt 0.5\tevcc$, we find that couplings
$\lambda^{\prime}_{2j1} \gtrsim 0.05$ and $\lambda^{\prime}_{21k} 
\gtrsim 0.03$ could be detected at the 200-GeV $\times$ 1-TeV collider at a 
luminosity of $1\fb^{-1}$.  This represents a considerable improvement 
in sensitivity over existing constraints on these couplings, as well 
as a significant improvement over the \textsc{hera} bounds on first-generation 
\rpv\ couplings.

\acknowledgments
We thank Heidi Schellman for helpful advice about angular cuts and 
trigger requirements.  C.Q. thanks the CERN Theory Division for warm 
hospitality during the summers of 1998 and 1999.

Fermilab is operated by Universities Research 
Association, Inc., under contract DE-AC02-76CHO3000 with the United 
States Department of Energy.

\begin{table}[tb]
\caption{Experimental constraints (at one or two standard deviations) on 
the \rpar--violating Yukawa couplings of interest, for the case of 
$200 \gevcc$ sfermions.  For arbitrary sfermion mass, multiply 
the limits by $(m_{\tilde{f}}/200\gevcc)$, except for 
$\lambda^{\prime}_{221}$.}
\begin{center}
\begin{tabular}{cc}
	%\hline
	\rpv\ Coupling & Limited by  \\
\hline
$\lambda^{\prime}_{21k} < 0.18~(1\sigma)$ & $\pi$ decay  \\
%\hline
$\lambda^{\prime}_{221} < 0.36~(1\sigma)$ & $D$ decay  \\
%\hline
$\lambda^{\prime}_{231} < 0.44~(2\sigma)$ & $\nu_{\mu}$ deep 
			inelastic scattering  \\
%	\hline
\end{tabular}
	\end{center}  
%\vspace*{24pt}
	\label{tab:lambdas}
\end{table}

\end{document}